\documentclass{PoS}
\usepackage{amsmath}
\usepackage{graphicx}
\usepackage{tikz}
\usepackage{slashed}

\title{SMOM - $\overline{\mbox{MS}}$ Matching for $B_K$ at Two-loop Order}

\ShortTitle{SMOM - $\overline{\mbox{MS}}$ Matching for $B_K$ at Two-loop Order}

\author{\speaker{Sandra Kvedarait\.e}\\
        Department of Physics and Astronomy, University of Sussex, Brighton BN1 9QH, UK\\
        E-mail: \email{S.Kvedaraite@sussex.ac.uk}}

\author{Sebastian J\"ager\\
        Department of Physics and Astronomy, University of Sussex, Brighton BN1 9QH, UK\\
        E-mail: \email{S.Jaeger@sussex.ac.uk}}

\abstract{The Kaon bag parameter, $B_K$, is a key non-perturbative ingredient in the search for new physics through CP-violation. It parameterizes the QCD hadronic matrix element of the effective weak $\Delta S=2$ four-quark operator which can only be computed non-perturbatively on the lattice. The perturbative matching of $B_K$ between the lattice renormalization schemes and $\overline{\mbox{MS}}$ scheme has been done before at one-loop order. In this proceedings we describe an ongoing calculation of the conversion factors for $B_K$ between the four non-exceptional RI-SMOM schemes and the $\overline{\mbox{MS}}$ scheme at two-loop order in perturbation theory.}

\FullConference{The 36th Annual International Symposium on Lattice Field Theory - LATTICE2018\\
		22-28 July, 2018\\
		Michigan State University, East Lansing, Michigan, USA.}

\begin{document}

\section{Introduction}

Indirect CP violation in the neutral kaon system $\epsilon_K$ is of pivotal importance in constraining new physics models. It is defined as a ratio 
\begin{equation}
\epsilon_K=\frac{A(K_L\rightarrow(\pi\pi)_{I=0})}{A(K_S\rightarrow(\pi\pi)_{I=0})},
\end{equation}
between the decay amplitudes of the long and short lived kaon to the two pion state of isospin 0. The Kaon bag parameter $B_K$ enters a dominant short distance contribution to $\epsilon_K$ and is given by
\begin{equation}
B_K=\frac{\langle K^0|Q_{VV+AA}|\bar{K}^0\rangle}{\frac{8}{3}f_K^2M_K^2}
\end{equation}
where $M_K$ is the kaon mass and $f_K$ is the leptonic decay constant \cite{BK}. $B_K$ parameterizes the QCD hadronic matrix element of the effective weak $\Delta S=2$ four-quark operator
\begin{equation}
Q_{VV+AA}=(\bar{s}\gamma_\mu d)(\bar{s}\gamma_\mu d)+(\bar{s}\gamma_5\gamma_\mu d)(\bar{s}\gamma_5\gamma_\mu d),
\end{equation}
which is non-perturbative and can be computed from first principles on the lattice.

Perturbative calculations for Wilson coefficients and anomalous dimensions are usually done in the $\overline{\mbox{MS}}$ NDR scheme. However, dimensional regularization can not be employed on the lattice. Instead, one possibility is a renormalization invariant momentum space subtraction (RI-SMOM) scheme. Hence, matching between the two schemes has to be performed. 

The proceedings is organised as follows. In Section 2 we define the RI-SMOM schemes and the conversion factors for the conversion to the NDR scheme. In Section 3 we describe the two-loop calculation. Section 4 contains details on the choice of the evanescent operators.

\section{RI-SMOM Schemes}

\begin{figure}[t!]
	\centering
	\includegraphics[trim=3.1cm 14cm 12cm 12cm, clip=true, width=6cm]{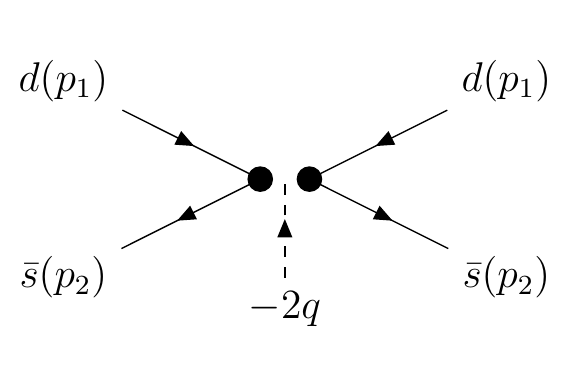}
	\caption{Non-exceptional kinematics for the four-quark operator: $p_1^2=p_2^2=(p_1-p_2)^2=p^2$ and $q=p_1-p_2$.}
\end{figure}

In order to renormalize the four-quark operator we use the same procedure for defining the kinematics of the momentum subtraction schemes as in the one-loop calculation \cite{BK}. RI-SMOM schemes are defined with non-exceptional kinematics, shown in Figure 1. The renormalization condition is defined such that at the subtraction point the four-point amputated Green's function $\Lambda$ is equal to the tree-level Green's function. The projection operators can be written as 
\begin{align}
P^{ij,kl}_{(1),\alpha\beta,\gamma\delta}&=\frac{1}{256N_c(N_c+1)}[(\gamma^\nu)_{\beta\alpha}(\gamma_\nu)_{\delta\gamma}+(\gamma^\nu\gamma^5)_{\beta\alpha}(\gamma_\nu\gamma^5)_{\delta\gamma}]\delta_{ij}\delta_{kl},\\
P^{ij,kl}_{(2),\alpha\beta,\gamma\delta}&=\frac{1}{64q^2N_c(N_c+1)}[(\slashed{q})_{\beta\alpha}(\slashed{q})_{\delta\gamma}+(\slashed{q}\gamma^5)_{\beta\alpha}(\slashed{q}\gamma^5)_{\delta\gamma}]\delta_{ij}\delta_{kl},
\end{align}
where $N_c$ is the number of colors, $i,j,k,l$ color and $\alpha, \beta, \gamma, \delta$ spinor indices. These projection operators are normalised in such a way that $Tr(P \Lambda^{tree})=1$. Having multiple projection operators allows us to assess the systematic uncertainties resulting from the choice of scheme.

The conversion factors can be defined as the ratio between the four-quark operator in the NDR scheme and in the RI-SMOM scheme given by 
\begin{equation}
Q_{VV+AA}^{NDR}(\mu)=C_{B_K}^{SMOM}(p^2/\mu^2)Q_{VV+AA}^{SMOM}(p)
\end{equation}
where $p$ is the renormalization scale of the SMOM scheme and $\mu$ is renormalization scale of the NDR scheme. We can compute the conversion factors using
\begin{equation}
C_{B_K}^{(X,Y)}=(C^{(Y)}_q)^2 P^{ij,kl}_{(X)\alpha\beta,\gamma\delta}\Lambda^{ij,kl}_{\alpha\beta,\gamma\delta},
\end{equation}
where $C_q$ is the conversion factor for the wave-function renormalization, $\Lambda^{ij,kl}_{\alpha\beta,\gamma\delta}$ is the amputated four-point Green's function computed in the $\overline{\mbox{MS}}$-NDR renormalization and at the RI-SMOM point. ($X,Y$) correspond to different RI-SMOM schemes.

\section{Two-loop Calculation}

\begin{figure}[b!]
	\centering
	\includegraphics[trim=3cm 23.5cm 4cm 3.6cm, clip=true, width=15.6cm]{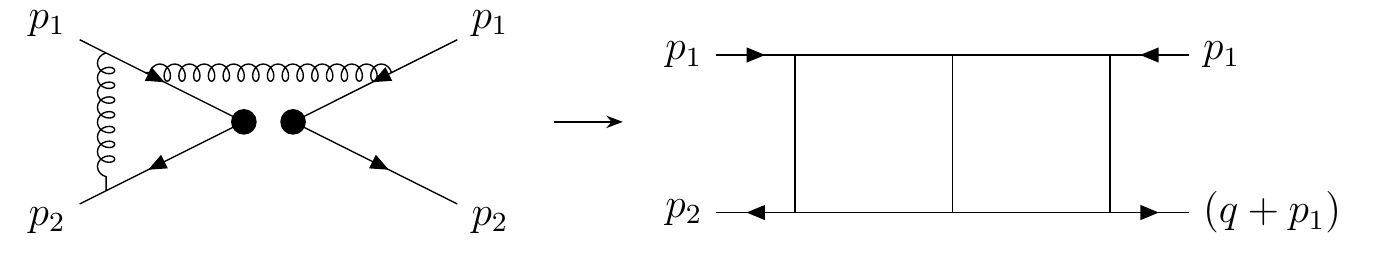}
	\caption{Obtaining an auxiliary topology based on one of the diagrams.}
\end{figure}

At two-loop order we have 28 independent diagrams \cite{Buras:1989xd}. Each of these diagrams can have up to six propagators which results in a large number of two-loop tensor integrals. The dimensional shift method can be used to reduce tensor integrals into scalar integrals \cite{Anastasiou:1999bn}. However, this would result in integrals with shifted dimensions. Instead of going through this process we decided to avoid doing tensor reduction by first contracting the diagrams with the projection operators and then doing the trace. This turns tensors into scalar products which can be expressed in terms of the denominators. 

We have to be careful about tracing over $\gamma_5$ in $D$ dimensions as traces such as $Tr(\slashed{p}_1\slashed{p}_2\slashed{k}_1\slashed{k}_2\gamma_5)$ can not be defined in a consistent way while requiring an anticommuting $\gamma_5$ \cite{Breitenlohner:1977hr}. However, when we perform such trace we find that the longest irreducible Dirac structure remaining after the tensor reduction is $Tr(\slashed{p}_1\slashed{p}_2\gamma_5)$ which is consistently equal to zero in 4 and in $D$ dimensions. Hence, we can anticommute all $\gamma_5$ to the right and drop the terms containing it without introducing ambiguities.

A two-loop diagram with two independent external momenta can have at most 7 linearly independent propagators. Hence, in order to systematically reduce the large number of integrals into a smaller set of master integrals we first have to define an auxiliarly topology. As shown in Figure 2, we do this by adding extra linearly independent propagators to one of the existing diagrams. Some diagrams may contain linearly dependent propagators from the start. By using the linear relationships between the propagators we can resolve this problem.

We find that we can map all 28 diagrams onto 5 different auxiliary topologies. We can use the integration by parts (IBP) identities \cite{Chetyrkin:1981qh} to express every integral that belongs to these topologies in terms of a small set of master integrals, given in Figure 3. If we used IBP reduction directly on the integrals without defining auxiliarly topologies we might run into problems as IBP's could give linear relationships between integrals with "extra" propagators. 

\begin{figure}[t!]
	\centering
	\includegraphics[trim=3cm 12cm 4cm 10.5cm, clip=true, width=15.6cm]{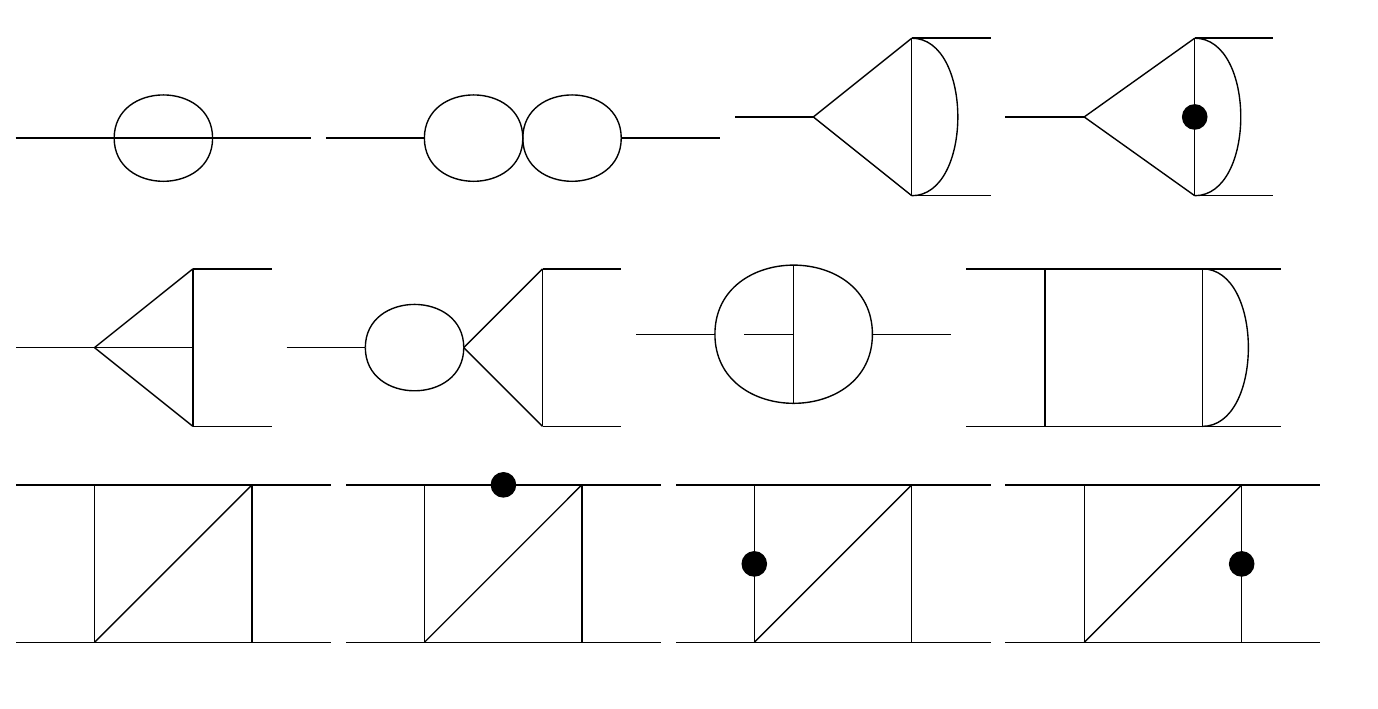}
	\caption{Two-loop master integrals. Black dots represent squared propagators.}
\end{figure}

All of the bubble and triangle diagrams have been calculated analytically in the literature \cite{Usyukina:1994iw}. However, there have been no analytic results for the box diagrams with four off-shell legs yet. We can calculate the box diagrams numerically using the sector decomposition method. Hence, the benefit of using the IBP method for reduction of a large number of integrals into a small set of masters is that it decreases the numerical uncertainties in the final result.

\section{Evanescent Operators}

The four-quark operator $Q$ is the only Kaon mixing operator in the Standard Model in 4 dimensions. However, in $D$ dimensions this is not necessarily the case. In $D$ dimensions we can also have a set of evanescent operators that vanish when taking $D\rightarrow4$. Our choice of evanescent operators is given by
\begin{align}
E_F&=(\bar{s}^k\gamma^\mu P_L d^j)(\bar{s}^i\gamma_\mu P_L d^l) - Q,\\
E_1^{(1)}&=(\bar{s}^i\gamma^{\mu_1\mu_2\mu_3}P_L d^j)(\bar{s}^k\gamma_{\mu_1\mu_2\mu_3}P_L d^l)-(16-4\epsilon-4\epsilon^2)Q,\\
E_2^{(1)}&=(\bar{s}^k\gamma^{\mu_1\mu_2\mu_3}P_L d^j)(\bar{s}^i\gamma_{\mu_1\mu_2\mu_3}P_L d^l)-(16-4\epsilon-4\epsilon^2)(Q+E_F),\\
E_1^{(2)}&=(\bar{s}^i\gamma^{\mu_1\mu_2\mu_3\mu_4\mu_5}P_L d^j)(\bar{s}^k\gamma_{\mu_1\mu_2\mu_3\mu_4\mu_5}P_L d^l)-(256-224\epsilon-144\epsilon^2)Q,\\
E_2^{(2)}&=(\bar{s}^k\gamma^{\mu_1\mu_2\mu_3\mu_4\mu_5}P_L d^j)(\bar{s}^i\gamma_{\mu_1\mu_2\mu_3\mu_4\mu_5}P_L d^l)-(256-224\epsilon-144\epsilon^2)(Q+E_F),
\end{align}
where $\gamma^{\mu_1\mu_2\mu_3}$ denotes the product of gamma matrices $\gamma^{\mu_1}\gamma^{\mu_2}\gamma^{\mu_3}$. 

This choice of evanescent operators allows us to use "Greek projections" of the form
\begin{equation}
P^{\alpha\beta,\gamma\delta}\Gamma^1_{\beta\gamma}\Gamma^2_{\delta\alpha}=Tr(\Gamma^1\gamma^\tau\Gamma^2\gamma_\tau),
\end{equation} 
to project out the evanescent part of the result \cite{Buras:1989xd}, \cite{Tracas:1982gp}.

Our choice of evanescent operators differs slightly from the ones used by J. Brod and M. Gorbahn \cite{Brod:2010mj} for the Wilson coefficients. The resulting change of basis can be obtained in terms of $1/\epsilon^2$ parts of renormalisation constants. The final result is the conversion factor from SMOM to NDR-$\overline{\mbox{MS}}$ a la Brod-Gorbahn for which the NNLO Wilson coefficients and anomalous dimensions are known.

\section{Conclusion}

In this proceedings we have presented an approach for calculating the two-loop conversion factors for $B_K$ at two-loop order. We have shown that all of the two-loop integrals can be expressed in terms of a small set of master integrals which can be evaluated analytically or numerically. Furthermore, we can evaluate the amplitude term by term by contracting with the "Greek projection" and thus remove all contributions from the evanescent operators. This allows us to avoid doing tensor reduction and we also find that the traces involving $\gamma_5$ can be consistently set to zero, therefore they do not produce ambiguities in D dimensions. Finally, we can convert the result to the Brod-Gorbahn operator scheme for which the NNLO Wilson coefficients and anomalous dimensions are known. This improved precision in evaluation of $B_K$ will help reduce the theory uncertainty on $\epsilon_K$ and increase sensitivity to the new physics effects.

\bibliographystyle{JHEP}

\begin{thebibliography}{1}
	
	\bibitem{BK}
	Y.~Aoki {\em et~al.}, {\it {Continuum Limit of $B_K$ from 2+1 Flavor Domain
			Wall QCD}},  {\em Phys. Rev.} {\bf D84} (2011) 014503,
	[\href{http://xxx.lanl.gov/abs/1012.4178}{{\tt 1012.4178}}].
	
	\bibitem{Buras:1989xd}
	A.~J. Buras and P.~H. Weisz, {\it {QCD Nonleading Corrections to Weak Decays in
			Dimensional Regularization and 't Hooft-Veltman Schemes}},  {\em Nucl. Phys.}
	{\bf B333} (1990) 66--99.
	
	\bibitem{Anastasiou:1999bn}
	C.~Anastasiou, E.~W.~N. Glover, and C.~Oleari, {\it {The two-loop scalar and
			tensor pentabox graph with light-like legs}},  {\em Nucl. Phys.} {\bf B575}
	(2000) 416--436, [\href{http://xxx.lanl.gov/abs/hep-ph/9912251}{{\tt
			hep-ph/9912251}}]. [Erratum: Nucl. Phys.B585,763(2000)].
	
	\bibitem{Breitenlohner:1977hr}
	P.~Breitenlohner and D.~Maison, {\it {Dimensional Renormalization and the
			Action Principle}},  {\em Commun. Math. Phys.} {\bf 52} (1977) 11--38.
	
	\bibitem{Chetyrkin:1981qh}
	K.~G. Chetyrkin and F.~V. Tkachov, {\it {Integration by Parts: The Algorithm to
			Calculate beta Functions in 4 Loops}},  {\em Nucl. Phys.} {\bf B192} (1981)
	159--204.
	
	\bibitem{Usyukina:1994iw}
	N.~I. Usyukina and A.~I. Davydychev, {\it {New results for two loop off-shell
			three point diagrams}},  {\em Phys. Lett.} {\bf B332} (1994) 159--167,
	[\href{http://xxx.lanl.gov/abs/hep-ph/9402223}{{\tt hep-ph/9402223}}].
	
	\bibitem{Tracas:1982gp}
	N.~Tracas and N.~Vlachos, {\it {Two Loop Calculations in {QCD} and the $\Delta
			I = 1/2$ Rule in Nonleptonic Weak Decays}},  {\em Phys. Lett.} {\bf 115B}
	(1982) 419.
	
	\bibitem{Brod:2010mj}
	J.~Brod and M.~Gorbahn, {\it {$\epsilon_K$ at Next-to-Next-to-Leading Order:
			The Charm-Top-Quark Contribution}},  {\em Phys. Rev.} {\bf D82} (2010)
	094026, [\href{http://xxx.lanl.gov/abs/1007.0684}{{\tt 1007.0684}}].
	
\end{thebibliography}

\providecommand{\href}[2]{#2}\begingroup\raggedright\endgroup

\end{document}